\documentclass{aa}  
\usepackage{multirow}
\usepackage{graphicx}
\usepackage{txfonts}
\begin{document}

   \title{Ancient and present surface evolution processes in the Ash region of
comet 67P/Churyumov-Gerasimenko}

   \author{A. Bouquety
          \inst{1},
          L. Jorda\inst{1}, O. Groussin \inst{1}, A. Sejourné \inst{2}, S. Bouley \inst{2} and F. Costard \inst{2}
          }

   \institute{Aix Marseille Univ, CNRS, CNES, LAM, Marseille, France\\
              \email{axel.bouquety@lam.fr}
         \and
             GEOPS, Université Paris-Saclay, CNRS, Rue du Belvédère,
Bât. 504-509, 91405 Orsay, France\\
             }


 
  \abstract
   {}
  {The Rosetta mission  provided us with detailed data of the surface of the nucleus of comet 67P/Churyumov-Gerasimenko. 
In order to better understand the physical processes associated with the comet activity and the surface evolution of its nucleus, we performed a detailed comparative morphometrical analysis of two depressions located in the Ash region.}
   {To detect morphological temporal changes, we compared pre- and post-perihelion high-resolution (pixel scale of 0.07--1.75~m) OSIRIS images of the two depressions. We quantified the changes using the dynamic heights and the
   gravitational slopes calculated from the Digital Terrain Model (DTM) of the studied area.
In particular, we measured seven geometric parameters associated with the two depressions (length, three width values, height, area, and volume) using the ArcGIS software before and after perihelion.}
   {Our comparative morphometrical analysis allowed us to detect and quantify the temporal changes 
that occurred in two depressions of the Ash region during the last perihelion passage.
We find that the two depressions grew by several meters.
The area of the smallest depression (structure I) increased by $90\pm20$\%, with two preferential growths: one close to the cliff associated with the apparition of new boulders at its foot, and a second one on the opposite side of the cliff.
The largest depression (structure II) grew in all directions,  increasing in  area by $20\pm5$\%, and no new deposits have been detected.
We interpreted these two depression changes as being driven by the sublimation of ices, which explains their global growth and which can also trigger landslides. 
The deposits associated with depression II reveal a stair-like topography, indicating that they have accumulated during several successive landslides from different perihelion passages.
Overall, these observations bring additional evidence of complex active processes and reshaping events occurring on short timescales (months to years), such as
depression growth and landslides, and  on longer timescales (decades to millenniums), such as cliff retreat.}
   {}


\titlerunning{Ancient and present surface evolution processes in the Ash region of comet 67P}
\authorrunning{Bouquety et al.}

\maketitle
%

\section{Introduction}

From August 2014 to September 2016 the Rosetta mission of the European Space Agency (ESA) allowed us to study in detail the comet 67P/Churyumov-Gerasimenko (hereafter 67P). 
In particular, the   Optical, Spectroscopic and Infrared
Remote Imaging System (OSIRIS; \citealp{Keller_2007_OSIRIS_CAM}) on board
Rosetta, returned high-resolution images of the nucleus
and its surface, revealing its bi-lobed structure \citep{Sierks_2015_nucleus_structure}. 

Geomorphological maps of the two
lobes have been performed by several authors, in particular \citet{Giacomini_2016_global_north_map}
for the northern hemisphere and \citet{Lee_2016_Geomorphological_mapping_south}
for the southern hemisphere. Due to the morphological diversity the nucleus has been divided into different regions and sub-regions, defined by their own morphological characteristics \citep{Thomas_2015_morphological_diversity,El-Maarry_2015_regional_morpho}. Using gravitational heights, gravitational
slopes,  size--frequency distribution of the boulders, morphological features, and/or colors, several regions have been studied in greater detail: Maftet, Ma'at, Nut, and Hatmehit \citep{Forgia_2015}; Abydos \citep{Lucchetti_2015_abydos};
Aswan \citep{Pajola_2016_aswan}; Imothep \citep{Auger_2015_Imhotep}; and Hapi \citep{Pajola_2019}.
The nucleus surface presents a wide variety of morphologies (\citealp{El-Maarry_2019}) with
dust-covered terrain (\citealp{Thomas_2015_Redistribution_of_particles}),
pits \citep{Thomas_2015_morphological_diversity}, boulders (\citealp{Pajola_2015}),
fractures at different scales (\citealp{El-Maarry_2015_regional_morpho,Matonti_2019_Bilobate_comet_morphology_shear_deformation}),
strata (\citealp{Massironi_2015_strata}), meter size thermal contraction cracks polygons (\citealp{Auger_2018_poly}),
and landslides (\citealp{Lucchetti_2019_landslide}).

\begin{figure*} [!ht]
\centering{\includegraphics[scale=0.35]{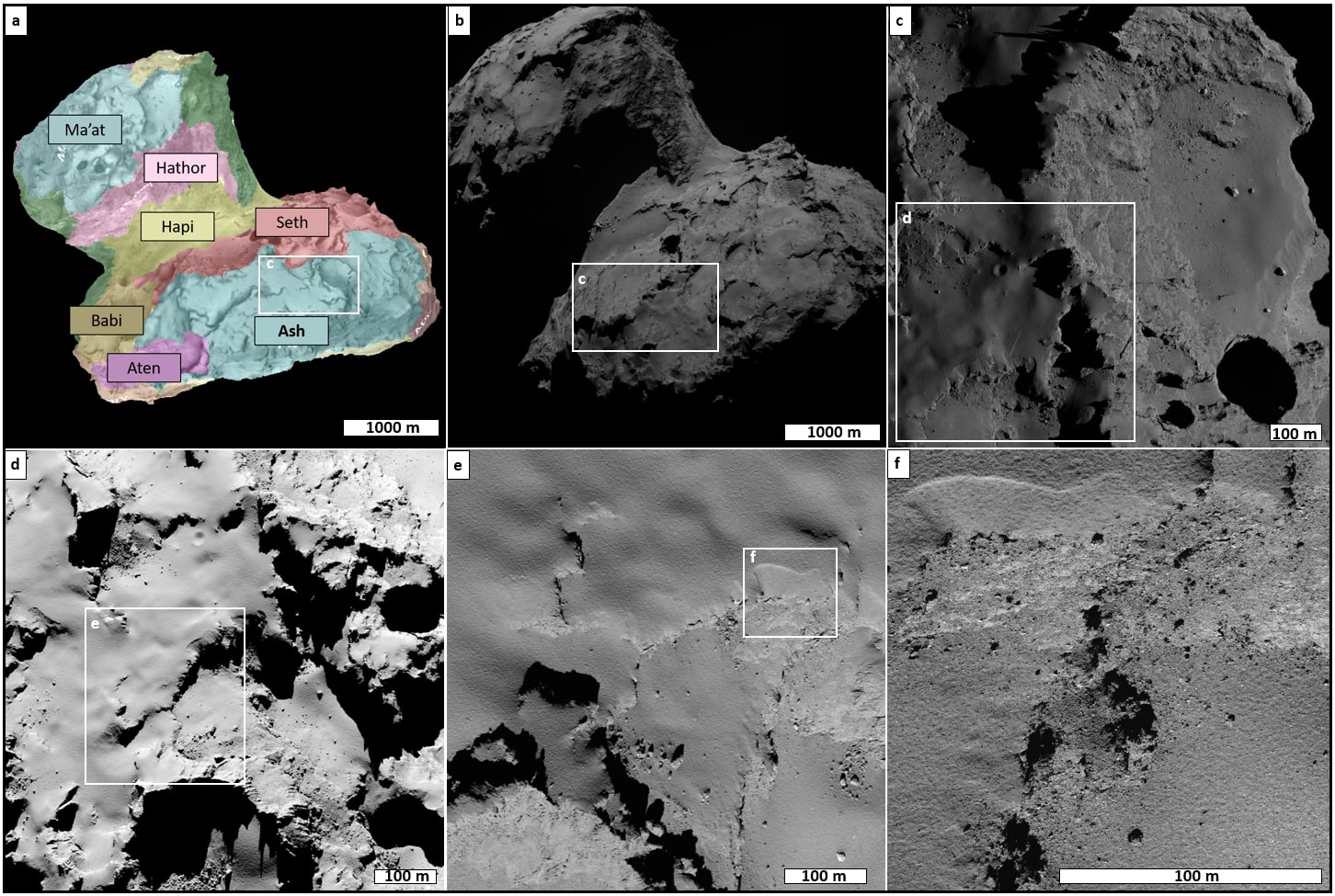}}
\caption{\label{fig.1} Ash region of the nucleus of comet 67P. The different images illustrate the different scales used in this study: (a) regions on the nucleus, (b) large-scale view  showing the bi-lobed nucleus, (c) global view of the Ash region, (d) studied area in Ash region, (e and f) magnified view of the studied area. The images are listed in Table \ref{tab:List-of-NAC}.}
\end{figure*}

In addition, several
studies have compared the OSIRIS images before and after perihelion in
order to better understand and constrain the surface evolution processes. From
December 2014 to June 2016, tens of morphological changes have been
observed. Changes arise as a modification or a creation of new landscapes
on the surface;  in particular, cliff collapse  appears to be a key process that reshapes the surface
\citep{Britt_2004,Pajola_2015,Pajola_2016_aswan,Pajola_2017,Steckloff_2016,Steckloff_2018,Lucchetti_2019_landslide}. In the Imhotep region
\citet{Groussin_2015_Geomorpho_Imhotep} observed morphological changes
in the form of roundish features growing in a preferential direction.
\citet{El-Maarry_2017} observed changes in several regions, such as cliff collapse in the Seth region, fracture
extension, or movement of decametric boulders. Most of those erosion or transient events have been interpreted as being linked to the seasonal and diurnal thermal 
cycle of the comet, both of which lead to ice sublimation and thermal fracturing 
\citep{Groussin_2015_Geomorpho_Imhotep,Pajola_2016_aswan,El-Maarry_2017,Hu_2017}. 

In this study  we continue the investigation
of 67P surface changes, focusing on the Ash region (Fig.~\ref{fig.1}). Located
on the largest lobe (i.e., the body),
the Ash region presents a wide variety of morphologies, and it is
a transitional area between the body and the neck (i.e., the part joining the two lobes). 
Activity has been detected in the Ash region, but only at a moderate level compared to more active regions, such as Hapi or Imhotep \citep{Lai_2019_seasonal_variation}, implying that the Ash region has not been entirely resurfaced during the Rosetta mission. It is therefore an ideal region to study its present and past surface dynamics.

The aim of this paper is to use a morphometrical approach to investigate
the Ash region. We focus on two areas within the Ash region, studying them before
and after perihelion to track surface changes.
We perform this analysis in order to constrain, quantify, and better understand
the landscape dynamic and evolution of the Ash region, 
aiming to find a generic eroding process that can apply to other nucleus regions. 
First, we describe the OSIRIS data set and the morphometrical approach we developed. 
Next, we present the results of our morphometrical analysis, followed
by a pre- and post-perihelion comparison to discuss the observed changes. 
Finally, we propose a qualitative scenario for the dynamical evolution of the studied area.

\section{Data and method: Comparative morphometrical analysis (CMA) }

\begin{figure*}
\centering{\includegraphics[scale=0.52]{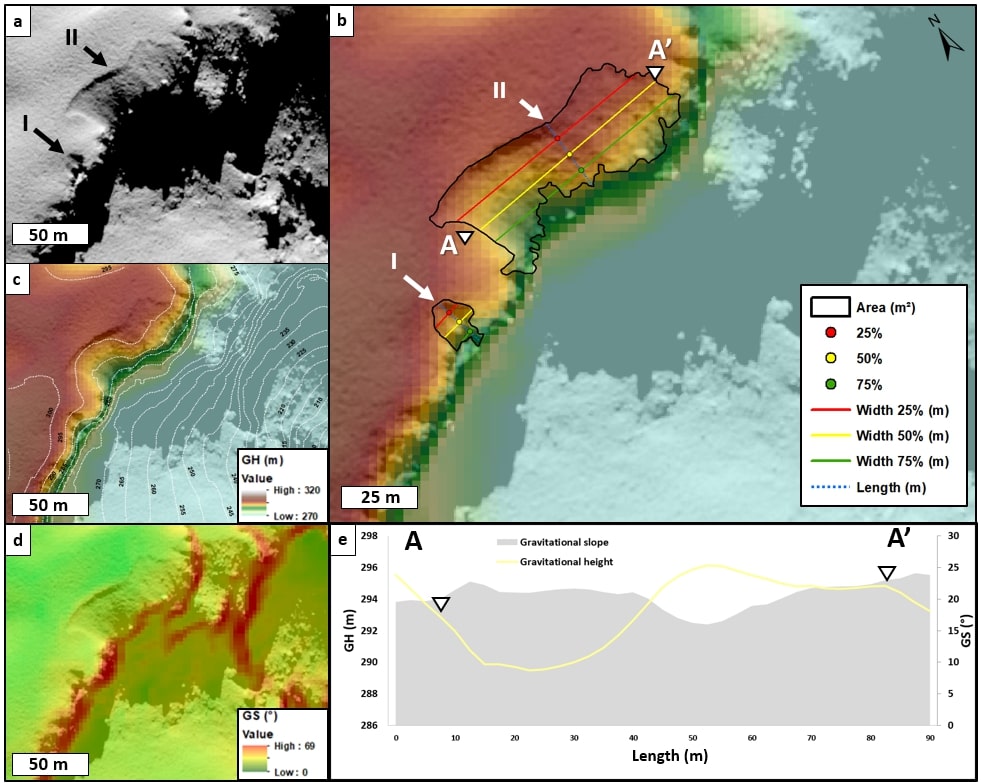}}
\caption{\label{fig:Studied-structures-and}Illustration of the  data products used for the morphometrical analysis (CMA):
(a) NAC image showing the two studied structures (I and II); (b) panel showing the length and width measurements with the gravitational height in background color (same color bar as panel c);
(c) gravitational height (GH) draped on the NAC image, with
5~m interval isolines in white; (d) gravitational slope (GS) draped on the NAC image; and (e) topographic and slope profiles of depression II, corresponding to the yellow line in panel b.}
\end{figure*}

The surface morphologies are the footprints left on the landscape by different
processes. Studying these footprints allows us to understand and
constrain the surface evolution processes that are affecting the body under study. 
The nucleus of 67P exhibits a wide range of morphologies \citep{Thomas_2015_morphological_diversity}
indicating that several processes occur at its surface. In
order to seek geomorphological evidence of these processes,
we performed a comparative morphometrical analysis (CMA), a technique 
successfully used to analyze other planetary surfaces (e.g., the Earth and Mars).
Based on \citet{Bouquety_2019,Bouquety_2020}, the CMA allows us to  study surface features 
via a morphological and geometrical approach, with a great level of detail, 
and to build a set of morphological parameters, which can then be
used for the pre- and post-perihelion comparison investigation. 
All our measurements were made with the ArcGIS software. 
The method is illustrated in Fig.~\ref{fig:Studied-structures-and}.

\subsection{Morphological analysis from the narrow-angle camera images}

The first step of the CMA is the morphological analysis using 
the high-resolution images of the OSIRIS instrument. We used seven images
of the narrow-angle camera, listed in Table~\ref{tab:List-of-NAC}: three images before perihelion (Nov. 2014 -- Feb. 2015), and four images
after perihelion (Jun. 2016 -- Sep. 2016). From these images, we identified two structures, each one on the top of a cliff, which seem to have been modified during the course of the mission (structures I and II; Fig.~\ref{fig:Studied-structures-and}a, b). In order to study these structures and their temporal changes, we defined the outline of each structure using three criteria:
(1) the edges must be continuous and easy to follow over tens of meters, (2) the texture inside
the edges must be different from that of the surrounding terrains (Fig.~\ref{fig:Studied-structures-and}a), 
and (3) the structure must be visible in at least three NAC images with different illumination conditions, to remove possible artifacts.

 \renewcommand{\arraystretch}{1.15}
\begin{table*}
\caption{\label{tab:List-of-NAC}List of OSIRIS narrow-angle camera images used in this study for both pre- and post-perihelion investigations. The spatial scale (m/px) is given for each image.}      
\centering          
\begin{tabular}  {l c c c c c} 
\hline\hline      
NAC image ID & Figure & Scale (m/pixel) & Emission  ($^{\circ}$) & Incidence ($^{\circ}$) & Phase  ($^{\circ}$)\\ 
\hline\hline
 & &Before Perihelion& \\
\hline
   N20150228T083347576ID4FF22 & Fig.1 (b) & 1.75 & 54.31 & 86.84 & 61.33 \\  
   \hline
   & Fig.1 (d), Fig.2, Fig.3 (a, c),& \\
   N20141110T140343285ID4BF61 & Fig.4, Fig.6 (a),Fig.7(a,c) & 0.55 & 52.43 & 69.66 & 73.03\\
    & Fig.8, Fig.9, Fig.11 (a)& & & & \\
    \hline
    
   N20150123T013527360ID4BF61 & Fig.6 (b), Fig.13 (a) & 0.54 & 30.29 & 71.99 & 93.31\\
   \hline
 & &After Perihelion& \\
\hline
   N20160618T013636681ID4EF22 & Fig.1 (c), Fig.10 & 0.54 & 46.72 & 100.37 & 67.08\\
\hline
    & Fig.1 (e), Fig.3 (b, d)&    \\
    N20160930T041121723ID4BF22 & Fig.5, Fig.7(b,d) Fig.11 & 0.24 & 23.65 & 61.58 & 41.61\\
    & Fig.14, Fig.15 (a)&\\  
\hline 
N20160906T005803924ID4FF41 & Fig.1 (f), Fig.15 (b) & 0.08 & 31.23 & 59.1& 89.12\\
\hline 
N20160906T011751583ID30F24 & Fig.15(c) & 0.07 & 40.11 & 54.49 & 93.11\\
\hline\hline 
    
\end{tabular}
\end{table*}

\subsection{Geometrical analysis}

The second step of the CMA is the quantitative geometric analysis of the selected
structures, which is performed using the digital terrain model (DTM).

\subsubsection{Digital terrain model}

\begin{figure*}
\centering{\includegraphics[scale=0.44]{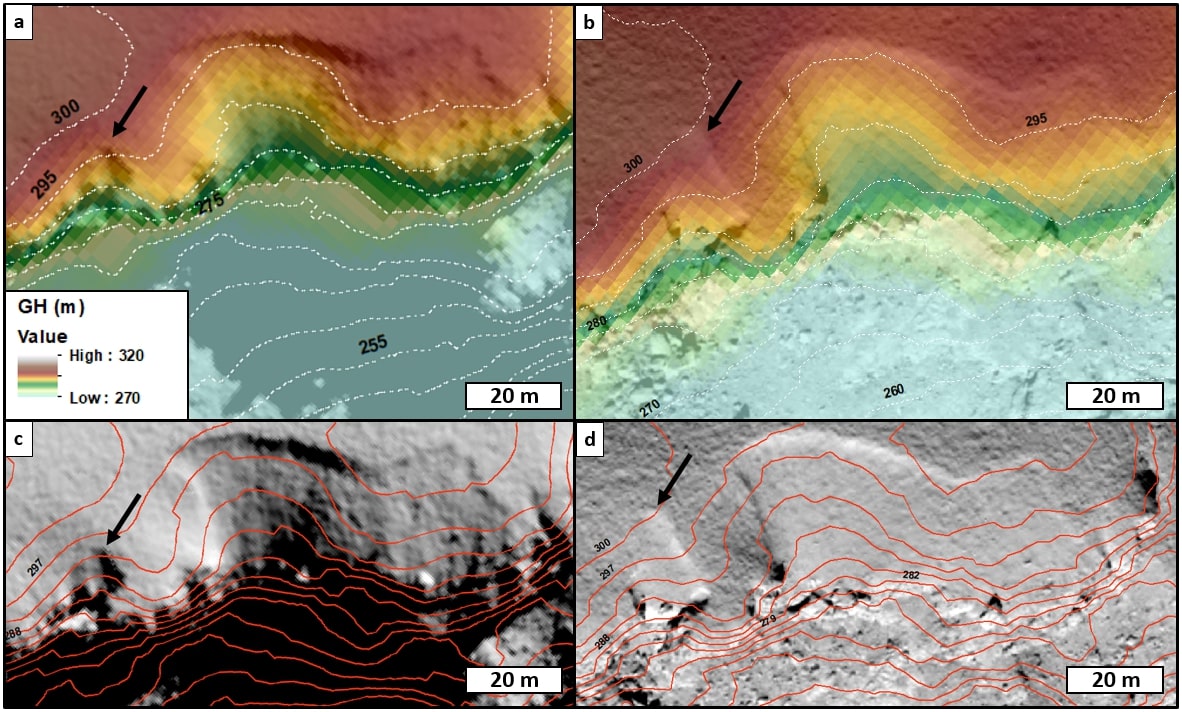}}

\caption{\label{fig:Images-versus-DTM}Comparison between the images before perihelion (a and c) or after perihelion (b and d) and the digital terrain model (DTM) overplotted on those images. In panels (a) and (b) the white dotted lines are 5~m interval isolines. In panels (c) and (d) the red lines are 3~m interval isolines. The black arrows indicate the main structure used for this comparison.}
\end{figure*}

The NAC data set offers stereo coverage of the Ash region,
used to derived a high-resolution digital terrain model (DTM) of this region via the method known as stereophotoclinometry  (SPC; Fig.~\ref{fig:Studied-structures-and}c, \ref{fig:Studied-structures-and}d; \citealt{GASKELL_2008,Capanna_2013,Jorda_2016}). 
However, since the SPC method mixes images from before and after perihelion, 
the resulting DTM represents a mean topography over the course of the Rosetta mission and does not track the surface changes that could have occurred at perihelion.
Depending on the date and number of images used to generate the DTM, it either better represents the surface before perihelion or after perihelion.

\renewcommand{\arraystretch}{1.15}
\begin{table*}
\caption{\label{tab:List-of-measured}List of measured parameters and method used for the measurement.}      
\centering          
\begin{tabular}  {l c c c} 
\hline\hline      
Measured parameters & Units & \multicolumn{2}{c}{Methods} \\ 

\multirow{2}{*} & \multirow{2}{*} &Before perihelion& After perihelion \\
\hline\hline 

   Length & m & Extracted from GH on ArcGIS & Calculated from pixel scale \\  
  
   Width & m & Extracted from GH on ArcGIS & Calculated from pixel scale \\
   
   Height & m & Calculated from topograpgic profile & Estimated from growth ratio\\

  Area & m$^2$  & \multicolumn{2}{c}{Calculated from pixel scale}\\
  Volume structure & m$^3$  & \multicolumn{2}{c}{Estimated from a triangular prisme shape
}\\
 Volume boulder & m$^3$  & \multicolumn{2}{c}{Estimated from a sphere geometry
}\\
\hline
    
\end{tabular}
\end{table*}

In our case we overlaid the DTM with
NAC images acquired before and after perihelion, using the ArcGIS software (Fig.~\ref{fig:Images-versus-DTM}),
to identify which ones provide the best fit to the Ash DTM.
Figures~\ref{fig:Images-versus-DTM}a and \ref{fig:Images-versus-DTM}c illustrate that 
structure I fits very well with the DTM before perihelion. The structure edge
follows the color limit and the isolines 295~m (Fig.~\ref{fig:Images-versus-DTM}a) and 294~m (Fig.~\ref{fig:Images-versus-DTM}c), which indicates that
the topographic structure observable in the NAC image is in agreement with the DTM. 
On the contrary, Figs.~\ref{fig:Images-versus-DTM}b and \ref{fig:Images-versus-DTM}d demonstrate that after perihelion the edges
of structure I do not follow the color lines or the isolines; moreover,
the edges cut the isolines 300~m (Fig.~\ref{fig:Images-versus-DTM}b) and 297~m (Fig.~\ref{fig:Images-versus-DTM}d) where a new depression
is now visible, while the isolines should skirt this structure. 
From the above comparison, we conclude that the DTM is a good reference only for the images acquired before perihelion.

To quantify the geometry of each structure before perihelion, we used the gravitational height (GH,
Figs.~\ref{fig:Studied-structures-and}c and \ref{fig:Studied-structures-and}e) and the gravitational
slope (GS, Figs.~\ref{fig:Studied-structures-and}d and \ref{fig:Studied-structures-and}e) derived from the DTM. For each structure, we measured seven parameters (length,
width (x3), area, height, and volume) and extracted four topographic
profiles and one slope profile (Table~\ref{tab:List-of-measured} and
Fig.~\ref{fig:Studied-structures-and}). 
For the structure after perihelion, we cannot use the DTM, and therefore we rely on the pixel scale of the images, as explained in Sect.~2.2.4.

\begin{figure*}
\centering{\includegraphics[scale=0.36]{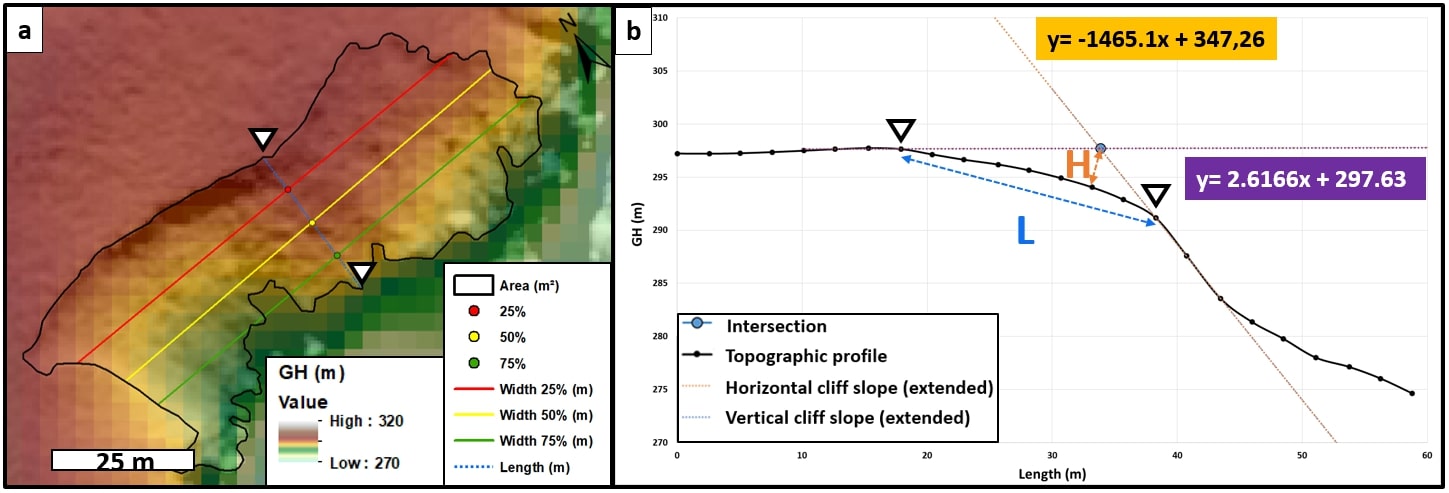}}

\caption{\label{fig:Example-of-volume}Illustration of the method used to calculate the volume: (a) length and width measurement from the images and (b) height measurement from the topographic profiles (see   text for details).}
\end{figure*}

\subsubsection{Length, width, and area}

The length of the structure is defined by the line
extending from the farthest upstream part, in the middle of the structure,
to the downstream part, orthogonally to the previously drawn contour
(Fig.~\ref{fig:Studied-structures-and}b).
To define the width we divided the length
into four segments of equal length. The width of the structure is then measured at each segment boundary, orthogonally to the length (i.e.,   three widths at 25\%, 50\%, and 75\% of the structure length; see Fig.~\ref{fig:Studied-structures-and}b). 
To allow a direct comparison between the pre- and post-perihelion parameters, we forced the length and width to be at the same position on the pre- and post-perihelion images.
For the length and for the three widths we can draw a topographic profile from the gravitational heights, and a slope profile from the gravitational slopes (Fig.~\ref{fig:Studied-structures-and}e).
The area of each structure, assumed to be planar, is extracted from the contour drawn in ArcGIS.

\subsubsection{Volume and height}

\begin{figure*} 
\centering{\includegraphics[scale=0.35]{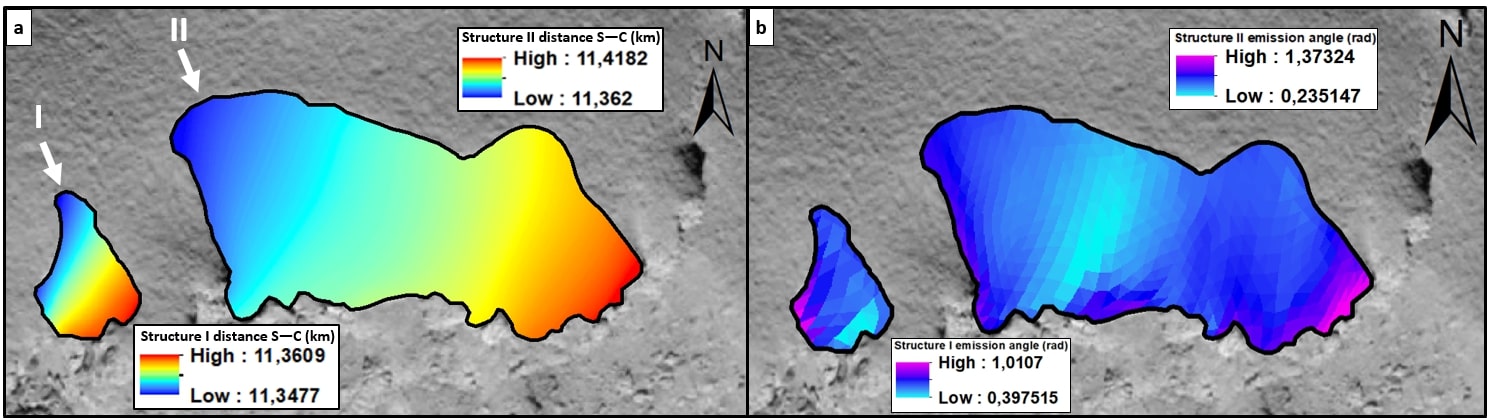}}

\caption{\label{fig:Example-of-pixel} Illustration of the pixel scale measurement from (a) the emission angle and (b) the distance between the camera and the surface (see Table~\ref{tab:Pixel scale calculation and pixel uncertainty} for details).}
\end{figure*}

\renewcommand{\arraystretch}{1.15}
\begin{table*}
\caption{\label{tab:Pixel scale calculation and pixel uncertainty}Pixel scale calculation and uncertainties.}      
\centering          
\begin{tabular}  {c c c c c c c c c} 
\hline\hline      
Parameters & Mean emission  & Mean distance  & Pixel scale & Pixel scale  &Pixel scale  & Pixel area & Pixel area  & Pixel area  \\ 
 &  angle (rad) &  S/C (m) &  (m/pixel)&   error (\%) &  error (m) &  (m$^2$/pixel)&   error (\%) &   error (m$^2$) \\ 

\hline\hline 

   Structure I & 0.60 & 11355 & 0.259 & 3 & 0.008 & 0.067 & 6 & 0.004 \\  
    Structure II & 0.57 & 11387 & 0.255 & 12 & 0.030 & 0.065 & 24 & 0.016 \\
\hline

\end{tabular}
\end{table*}

The volume of the structure is defined by the volume of material (in m$^3$) removed during the collapse (or landslide). To compute this volume we had to estimate the original shape of the structure before the collapse. 
We used a simple geometric approach and assumed a triangular prism shape for the removed material (Fig.~\ref{fig:Example-of-volume}b). The triangular prism is defined by its length $L$ and its width $W$, as explained in the previous section, and by its height $H$. To estimate the height $H$, we extrapolated the topographic profiles measured outside the structure on both sides, which correspond to the orange and purple lines in Fig.~\ref{fig:Example-of-volume}b, and defined $H$ as the distance between the surface and the intersection of these two lines. The volume $V$ can then be calculated as
\begin{equation}
 V = \frac{1}{2} L  W  H
.\end{equation}
We computed the length, width, and height at different positions along the structure, from which we derived the minimum volume $V_{min}$ and the maximum volume $V_{max}$ of each structure, using the minimum and maximum values of each parameter $L$, $W$, and $H$ (Table~\ref{tab:Value of each measured parameter}). 

\subsubsection{Pixel scale measurement and growth ratio}

As explained in section 2.2.1, the DTM is not consistent with post-perihelion images;
we can therefore only rely on the images themselves to constrain the geometry of the two structures after perihelion. 
For this purpose we must accurately estimate the pixel scale of the images (in~m/pix), that can be calculated from two different data sets. The  pointing direction and Spacecraft--Comet (S--C) coordinates  in body frame are derived from the stereo SPC solution of each image \citep{Jorda_2016}, and the SPC global shape model (at about 1.5~m sampling).
From these data it is possible to calculate the distance between the S--C (i.e., the camera) and the surface, as well as the mean emission angle
in each pixel (Fig.~\ref{fig:Example-of-pixel} and Table~\ref{tab:Pixel scale calculation and pixel uncertainty}).
We then calculate the mean distance $d$ and emission angle $e$ of each structure by averaging the values
for each pixel inside them.
The final pixel scale is calculated as $p \sim d \times \mathrm{IFOV} / \cos{e}$, where $\mathrm{IFOV} = 18.82$~microrad is 
the pixel field of view.
We obtain a value of $0.259\pm0.008$~m/pixel for structure I and $0.255\pm0.030$~m/pixel for structure II, where the uncertainty is estimated from the rms value of $d$ and $1 / \cos{e}$.

This method  partially overcomes the lack of DTM, but we cannot extract 
the topographic profiles to calculate the height, and therefore the volume.
We thus make the assumption that the change in height, or vertical plane, is proportional to the change in length and width, or horizontal plane (i.e., there is a global growth ratio for each structure).

In addition to the uncertainty coming from the pixel scale, which applies to all measurements, there is also the error on the measurements themselves. This error is $\pm$2~pixel for all length, width, and height measurements, and is equal to the perimeter (in pixels) for area measurements. We computed the overall error using a root sum squared of the pixel scale error and the measurement error. The growth ratio, however, is only characterized by the measurement error since the pre- and post-perihelion values are affected by the same pixel scale uncertainty that vanishes.

\begin{figure*} [!ht]
\centering{\includegraphics[scale=0.37]{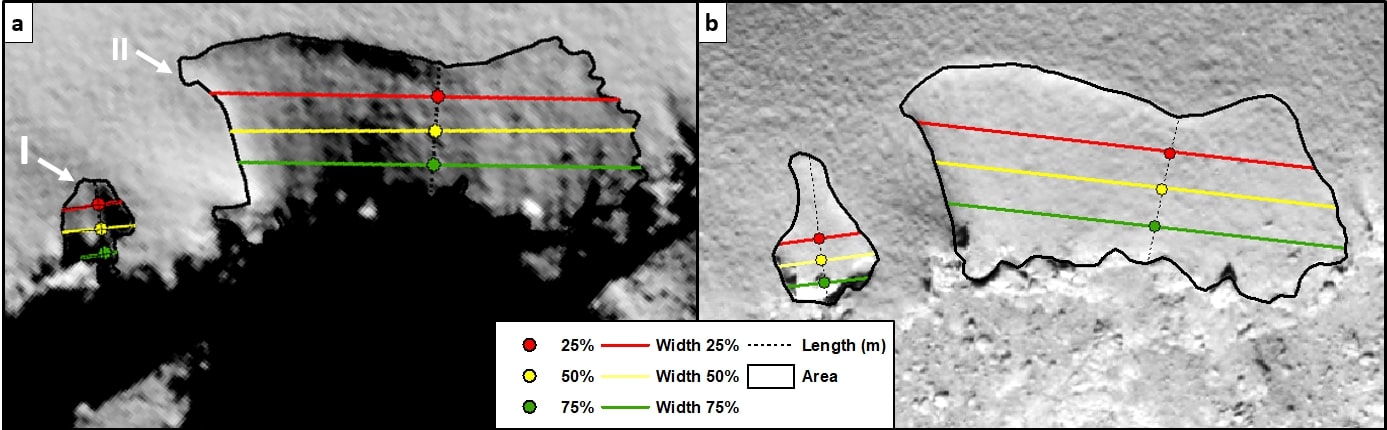}}
\caption{\label{fig8} Summary of all the geometric measurements made for depressions I and II, before and after perihelion.
(a) Image of depressions I and II (0.55 m/pixel), before perihelion, showing the measured length and widths. 
(b) Image of the two depressions (0.24 m/pixel), after perihelion, showing the measured length and widths. 
Table~\ref{tab:Value of each measured parameter} summarizes the value of each measured parameter.
}
\end{figure*}

\begin{table*}
\caption{\label{tab:Value of each measured parameter}Value of each measured parameter.}      
\centering          
\begin{tabular}  {c c c c c c c c c} 
\hline\hline      
Parameters & Length & Width & Width& Width & Area &  & {Volume (m$^3$)}  &\\ 
 &  (m)  &  25\% (m) &  50\% (m) &   75\% (m) &   (m$^2$) &  Min &   Max & Mean \\ 

\hline\hline
Before perihelion & & & & & & & &  \\
\hline
\multirow{2}{*}{Depression I} & 17.6  & 10.9  & 13.2 & 7.1 &195&200&855&540 \\  
    &  ($\pm \ 0.7$)&($\pm \ 0.6$)&($\pm \ 0.7$)&($\pm \ 0.6$)&($\pm \ 17$)&($\pm \ 9$)&($\pm \ 36$)&($\pm \ 23$)\\
\multirow{2}{*}{Depression II} & 25 & 76.1 & 75 & 75 & 2479 & 9370 & 17280 & 13320 \\
   &  ($\pm \ 3$)&($\pm \ 9.1$)&($\pm \ 9$)&($\pm \ 9$)&($\pm \ 598$)&($\pm \ 1590$)&($\pm \ 2932$)&($\pm \ 2290$)\\
\hline  
After perihelion& & & & & & & &\\
\hline
\multirow{2}{*}{Depression I} & 32.1  & 17.4  & 20.2 &17.21 & 376 &380&3105&1740 \\  
    &  ($\pm \ 1.1$)&($\pm \ 0.7$)&($\pm \ 0.8$)&($\pm \ 0.7$)&($\pm \ 32$)&($\pm \ 16$)&($\pm \ 132$)&($\pm \ 74$)\\
\multirow{2}{*}{Depression II} & 30.2 & 80.7 & 82.2 & 81 & 2848 & 11200 & 22700 & 16950 \\
   &  ($\pm \ 3.7$)&($\pm \ 9.7$)&($\pm \ 9.9$)&($\pm \ 9.7$)&($\pm \ 686$)&($\pm \ 1901$)&($\pm \ 3852$)&($\pm \ 2677$)\\
 \hline
    
\end{tabular}
\end{table*}

\section{Results of the morphometrical analysis}

We present the measurements of the morphometrical parameters defined in the previous section, 
for structures I and II, before and after perihelion. These morphological and geometric parameters  allow us to quantify these two structures, and to compare their evolution during the perihelion passage. All the measurements are summarized in Fig.~\ref{fig8} and Table~\ref{tab:Value of each measured parameter}.

\subsection{Before perihelion results}

\subsubsection{Morphological description of structures I and II}

\begin{figure*}
\centering{\includegraphics[scale=0.49]{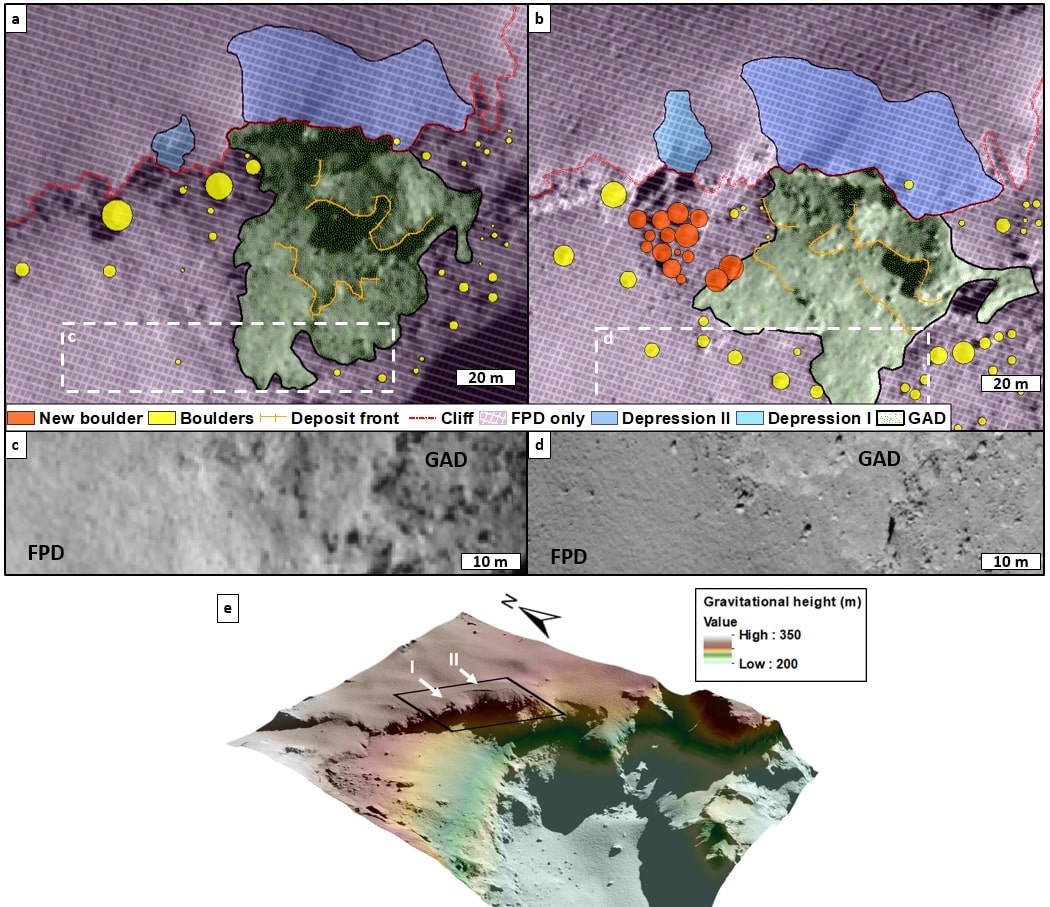}}
\caption{\label{fig:NAC-images-of}Studied area in Ash region. (a) and (b) Geomorphological map of the studied area before and after perihelion (0.54 m/pixel). The different shape of the gravitational accumulation deposits (GAD) is due to different viewing angles between the two images (see Table~\ref{tab:List-of-NAC}). (c) and (d) Magnified image illustrating the different texture of the fine particle deposit (FPD) compared to the GAD, before and after perihelion. (e) Digital terrain model of the studied area, with a NAC image overlaid, and gravitational heights in color.}
\end{figure*}

The studied area is located in the Ash region, near Aswan, and 
is composed of a cliff 30~m in height (GH). This cliff is the boundary
between two plateaus, the higher one with an elevation of 300~m (GH) and
the lower one at 270~m (Fig.~\ref{fig:NAC-images-of}). The studied
structures are both located on  cliff edges. They are in an area covered
by  material with an apparent smooth and homogeneous texture at the meter scale (Fig.~\ref{fig:Studied-structures-and}a
and Fig.~\ref{fig:NAC-images-of}), called fine particles deposit (FPD) following the geomorphological mapping criteria
established by \citet{Lee_2016_Geomorphological_mapping_south} and
\citet{Giacomini_2016_global_north_map}. The two structures are separated by only 25~m. They appear as two depressions, with well preserved and continuous edges, the cliff being their lower limit (Fig.~\ref{fig:NAC-images-of}). 

The two depressions do not have the same shape or the same dimension. Depression I, 
the smaller one, is about 20~m wide, and is shaped like 
a spade  where the lower part seems to notch the cliff
(Fig.~\ref{fig:NAC-images-of} and Fig.~\ref{fig:Morphometric-measrument-for}a).
Depression II, the larger one, is about 80~m wide, and has an irregular shape; it can be divided into two distinct parts, with a hook shape on
its western edge (see section 3.1.3 for more details). 

The two depressions are filled by material, most probably FPD. In depression II, we observe that the filled material appears to be different depending on the incidence angles of the images, revealing topographical variations inside this depression (Fig.~\ref{fig:Example-of-pixel}a; see also Table~\ref{tab:Pixel scale calculation and pixel uncertainty} and Fig.~\ref{fig:NAC-images-of}).

Finally, we observe
that the material at the foot of the cliff below depression II has a more granular and bulky texture, and is composed of boulders and indurated material that covers the surrounding FPD terrains  (Figs.~\ref{fig:NAC-images-of}c and d).

\subsubsection{Geometry of depression I}

\begin{figure*} 
\centering{\includegraphics[scale=0.60]{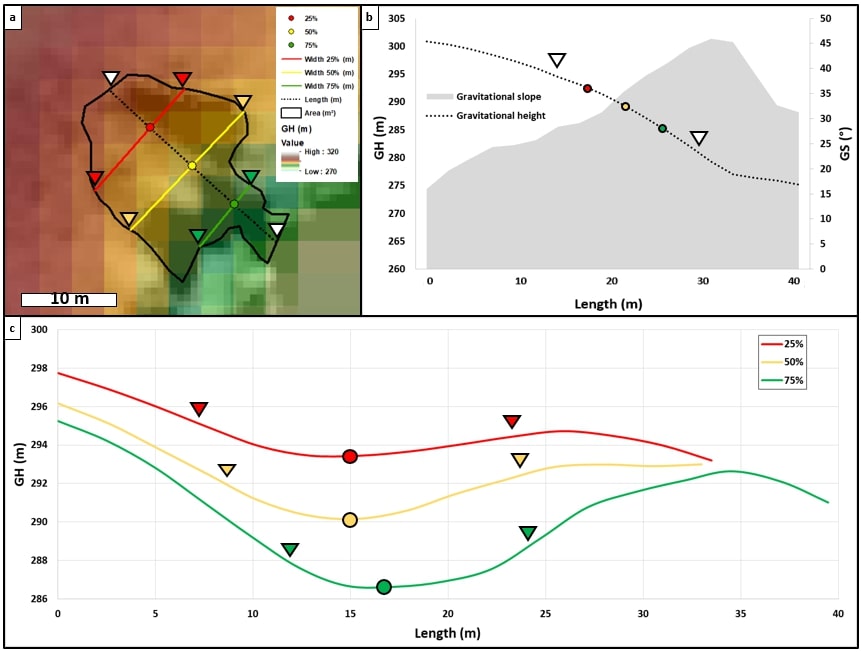}}

\caption{\label{fig:Morphometric-measrument-for} Illustration of the geometric measurements for depression I. 
(a) Morphometric measurements showing the length and widths.
(b) Topographic profile and slope profile along the length.
(c) Transverse topographic profile along the different widths, with a vertical exaggeration of 2.8.}
\end{figure*}

\begin{figure*}
\centering{\includegraphics[scale=0.39]{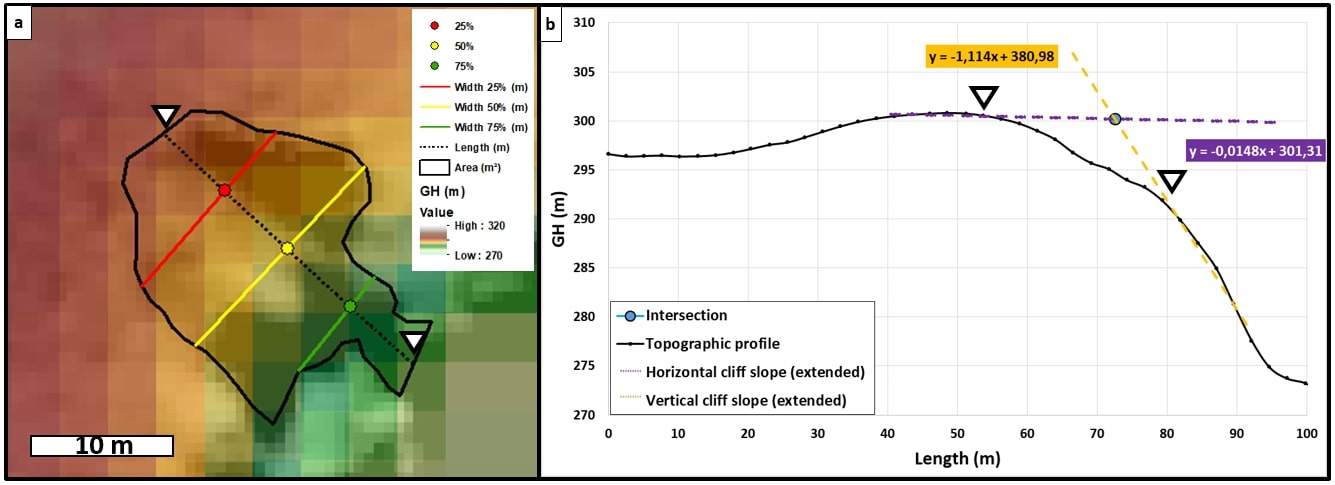}}
\caption{\label{fig:Length-profile-used}Length profile used to estimate the volume of depression I. 
(a) Location of the length profile in depression I. (b) Topographic profile along this length.}
\end{figure*}

In order to quantify the geometry of the depressions, we used the DTM calculated via the SPC method (see section  2.2.1).
Depression I is longer than it is wide, with a length of 17.6~m and a
mean width of 10.4~m (Fig.~\ref{fig:Morphometric-measrument-for}).
Upstream, in the first 7 meters, the slope increases slowly 
from 25$^\circ$ to 30$^\circ$. Closer to
the cliff the slope increases rapidly to 45$^\circ$ . The transverse
profiles indicate that the depression is shaped like a bowl  (Fig.~\ref{fig:Morphometric-measrument-for}c).
The depth of the depression seems to increase from 1~m depth upstream
to 4.5~m depth downstream. The depth and slope increase following the length line. The estimated collapsed volume
from depression I is between 200~m$^{3}$ (Vmin) and 885~m$^{3}$ (Vmax)
(Fig.~\ref{fig:Length-profile-used} and Table~\ref{tab:Value of each measured parameter}). 

\subsubsection{Geometry of depression II}

\begin{figure*}
\centering{\includegraphics[scale=0.67]{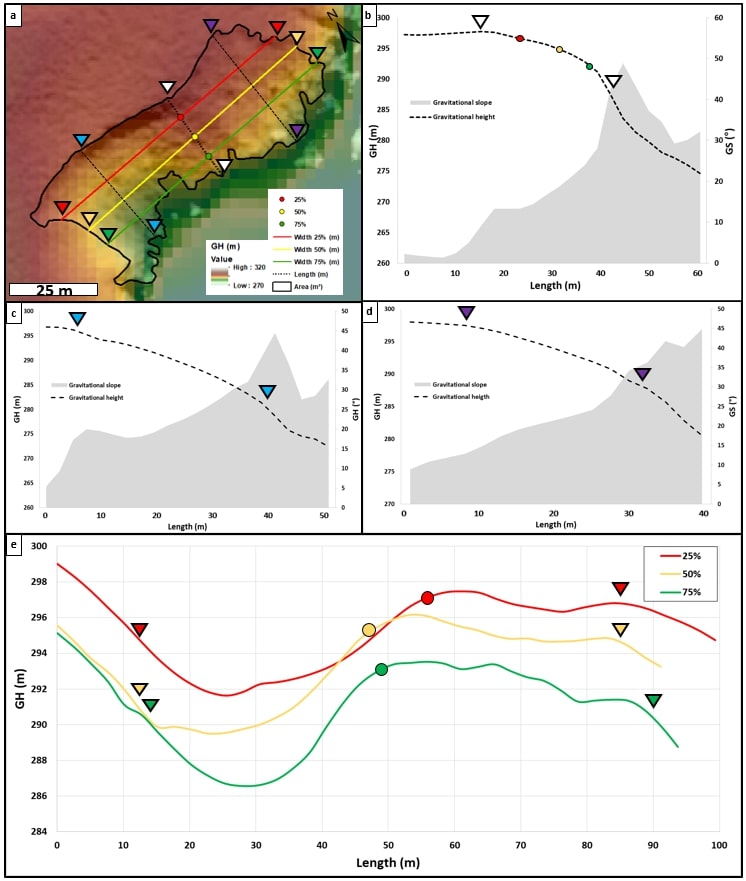}}

\caption{\label{fig:7}Illustration of the geometric measurements for depression II. 
(a) Morphometric measurements showing the length and widths.
(b, c, d) Topographic profile and slope profile along the three lengths.
(e) Transverse topographic profile along the different widths, with a vertical exaggeration of 6.7.}
\end{figure*}

Depression II is wider than longer, with a length that varies between
25~m and 35~m (Fig.~\ref{fig:7}a-d) and a mean width of 75.3~m (Fig.~\ref{fig:7}e).
The three length profiles show a similar pattern: in the first meters upstream the slope is around 18$^\circ$, and then it increases to 45$^\circ$ closer to the cliff (Fig.~\ref{fig:7}b-d). These values were extracted from the DTM.

The transverse profiles reveal that depression II has two distinct
parts (Fig.~\ref{fig:7}a and e). The first goes
 from the western edge to the middle of the depression. The topography shows a 5 to 6~m depth bowl shape, and the depth increases closer to the cliff. The second goes from the middle to the eastern edge; the topography is flatter and shallower (Fig.~\ref{fig:7}a and e).

\begin{figure*}
\centering{\includegraphics[scale=0.42]{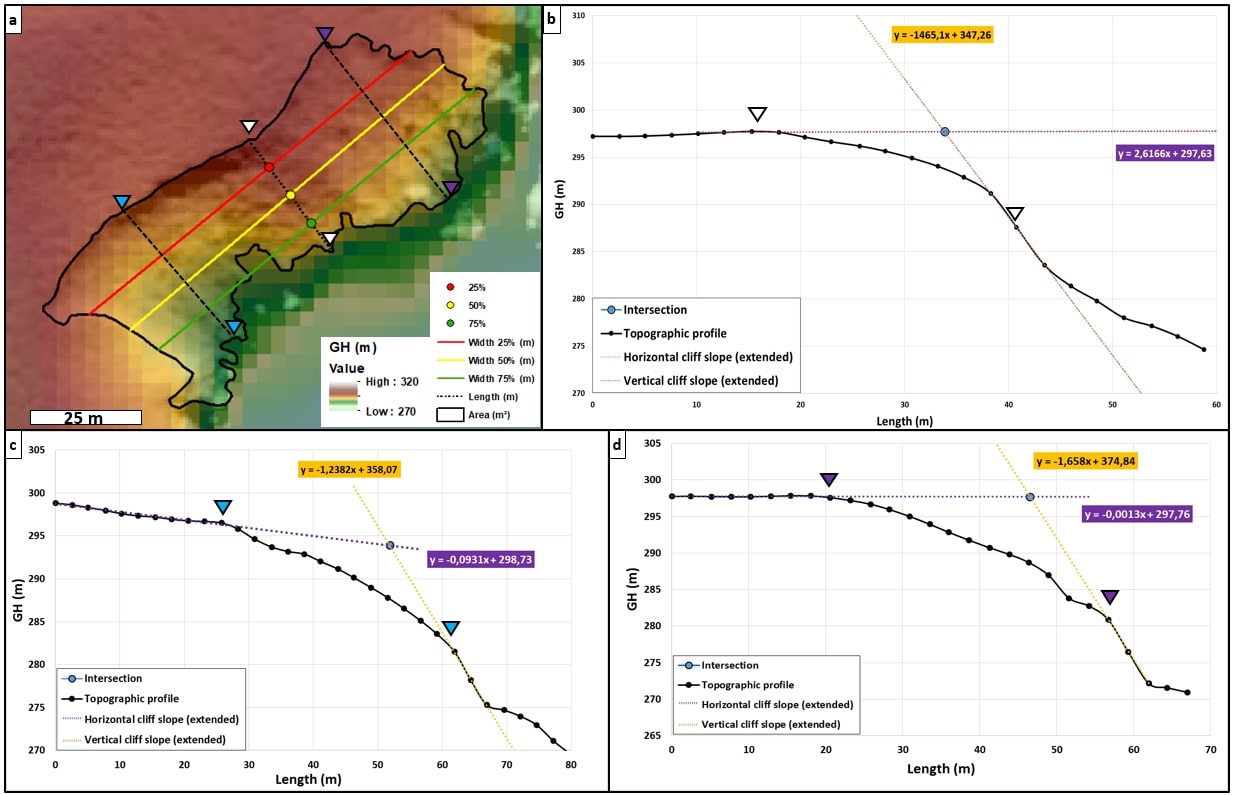}}
\caption{\label{fig:Length-profile-DII}Different length profiles used to estimate the volume of depression II. 
(a) Location of the three length profiles in depression II.
(b, c, d) Topographic profile along each length.}
\end{figure*}

The estimated collapsed volume
from depression II is between 9370~m$^{3}$ (Vmin) and 17280~m$^{3}$ (Vmax)
(Table~\ref{tab:Value of each measured parameter} and Fig.~\ref{fig:Length-profile-DII}). The large uncertainty in the volume results from the above variations of the transverse topographic profiles.

\subsection{After perihelion}

\subsubsection{Morphological description of  depressions I and II}

\begin{figure*}
\centering{\includegraphics[scale=0.5]{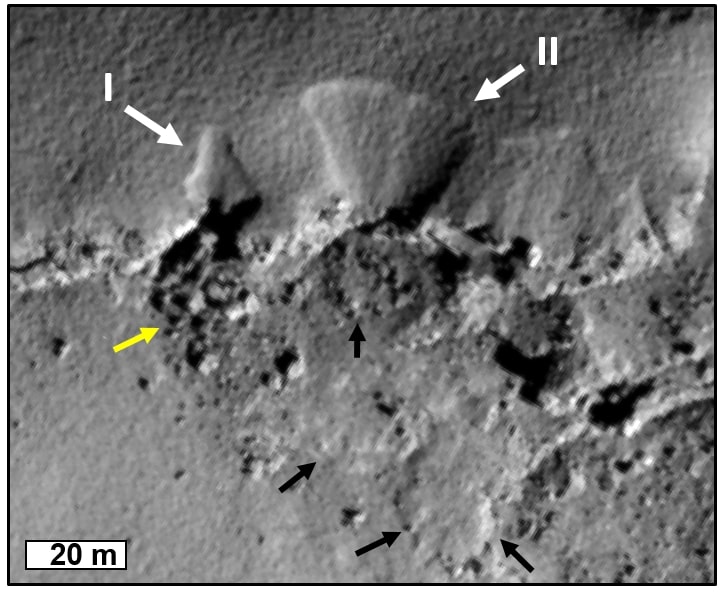}}

\caption{\label{fig:11}NAC image showing the two depressions after perihelion.
The white arrows indicate the two depressions I and II, the black arrows delimit the gravitational accumulation deposits, and the yellow arrow indicates the boulder's position. }
\end{figure*}

The two depressions and their associated deposits, observed after perihelion, are shown 
in Fig.~\ref{fig:11}. 

Depression I keeps
its spade shape, and is still longer than wider (section 3.2.2, Fig.~\ref{fig:11}). We observe ten-meter boulders at the foot of the  cliff, some of them located 30~m away from the cliff; these boulders were not present before perihelion (Fig.~\ref{fig:NAC-images-of}a and b). 

Depression II keeps its irregular shape. We observe material at the foot of the  cliff  that has  a granular texture that could be made of boulders covered by FPD, similar to the pre-perihelion observations. The boulders that compose this material are smaller, and seem more degraded than the boulders at the foot of depression I. The material at the foot of depression II extends far away from the cliff, up to 100~m downhill (Fig.~\ref{fig:NAC-images-of} and Fig.\ref{fig:11}).

\subsubsection{Geometry of depressions I and II}

Depression I is still longer ($\sim$31~m) than wider ($\sim$17~m), and depression II is still wider ($\sim$81~m) than longer ($\sim$30~m) (Table~\ref{tab:Value of each measured parameter}). As explained in section 2.3, the lack of DTM after perihelion did not allow us to study the topographic and slope profiles. We   nevertheless estimated the volume of depression I to be in the range 380--3105~m$^{3}$, and that of depression II to be in the range 11200--22700~m$^{3}$

Overall, the two depressions have increased in size compared to before perihelion. Based on the data set of Table~\ref{tab:Value of each measured parameter}, we   quantify and discuss these changes in section 4.

\section{Comparing the results of the morphometrical analysis before and after perihelion }

\subsection{Morphological comparison}

\begin{figure*} 
\centering{\includegraphics[scale=0.56]{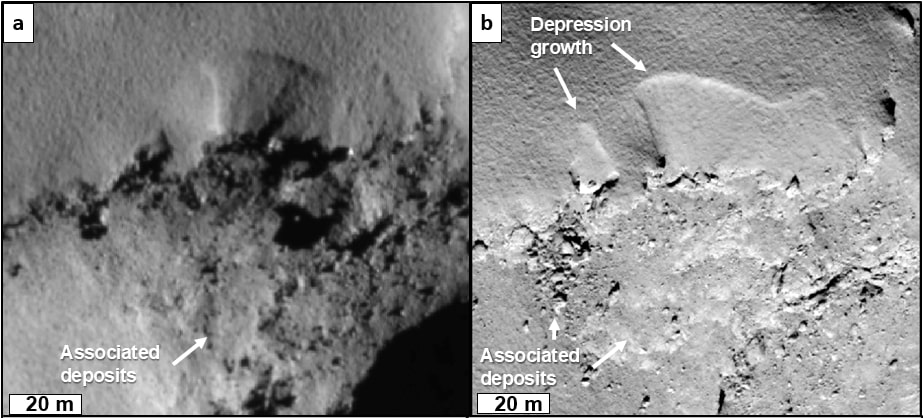}}
\caption{\label{fig:Morphological-comparison-of}Two images showing the depressions before (a) and after (b) perihelion. The spatial resolution is 0.54~m/pixel and 0.24~mm/pixel respectively. The white arrows indicate the main areas where the two depressions grew. See section 4 and Fig.~\ref{fig:NAC-images-of} for a detailed description of the comparison. Table~\ref{tab:Value of each measured parameter} provides the estimated value of the volume of deposits associated with each depression.}
\end{figure*}

The first step is to look for temporal morphological
changes in the two depressions and in their associated deposits.
Figure~\ref{fig:Morphological-comparison-of} shows a comparison
of the studied area using two NAC images acquired before and after perihelion
under   similar geometric conditions (distance, incidence, emission, and phase angles).
Visually, we identified three majors changes after perihelion, 
one in each depression and another   in the deposit associated with depression I: 
\begin{itemize}
 \item The shape of depression I has been modified; 
the western edge is more rounded than before perihelion. 
The length of depression I has increased, and it  is now as long as depression II.
 \item The edge of depression II has been modified. 
While the western edge had a hook shape before perihelion, it is now straighter 
and the hook  is milder than before perihelion. 
 \item New boulders appeared during the perihelion passage at the foot of the cliff associated with depression I. The size of these boulders is in the range 2.2--6.6~m. 
\end{itemize}

\begin{table}
\caption{\label{tab:Values of measured volume}Volume of the deposits for depressions I and II.}      
\centering          
\begin{tabular}  {c c c } 
\hline\hline      
 \multicolumn{2}{c}{Parameter}  &  \multirow{2}{*}{Volume (m$^3$)}\\ 
 Structure &  Period  &   \\ 
 
\hline\hline
\multirow{2}{*}{Depression I} & Before perihelion  & No observed deposits   \\  
    &  After perihelion & 430 ($\pm \ 34$)\\
\hline  
\multirow{2}{*}{Depression II} & Before perihelion  & 29900 ($\pm \ 7800$)\\  
    &  After perihelion & Unchanged \\
 \hline
    
\end{tabular}
\end{table}

These temporal morphological changes indicate that the cometary surface
was   active at perihelion, in agreement with previous works 
\citep{Groussin_2015_Geomorpho_Imhotep,Pajola_2016_aswan, Pajola_2017, El-Maarry_2017,Hu_2017}.

\subsection{Geometrical comparison }

\begin{figure*} 
\centering{\includegraphics[scale=0.52]{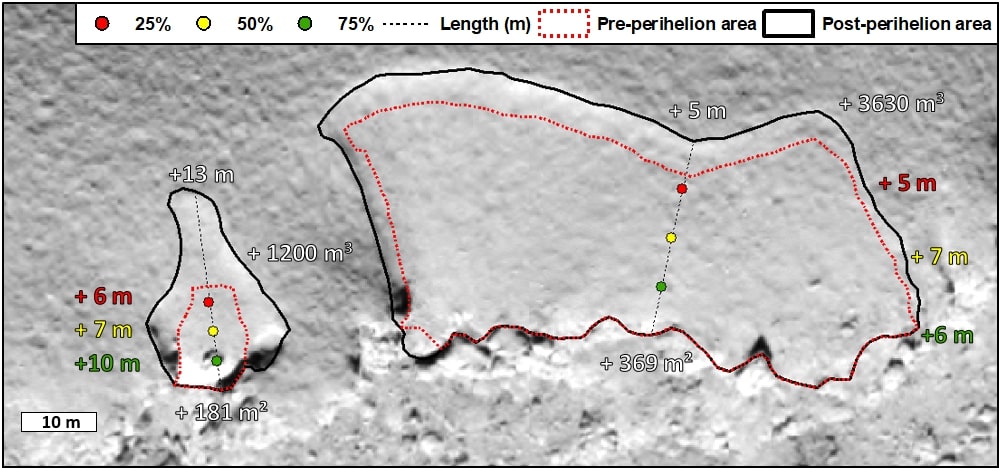}}
\caption{\label{fig:Geometrical-comparison-of}Geometrical comparison of the two depressions, showing the measured changes.
The dotted red lines indicates the shape of the depression before perihelion; there is an uncertainty of about 1~m on the position of this contour, due to the different illumination conditions of the images used to define the contour before and after perihelion. The increase in lengths, widths, and areas, are labeled.  Table~\ref{tab:Value of growth and growth ratio for each parameters} summarizes the measured changes, with the value of growth and growth ratio for each parameter.}
\end{figure*}

As suggested by the morphological comparison, the two depressions
grew after perihelion (Fig.~\ref{fig:Morphological-comparison-of}).
We used the geometrical parameters presented in section 3 (Fig.~\ref{fig8} and Table~\ref{tab:Value of each measured parameter}) to quantify these temporal
morphological changes.
From the pre- and post-perihelion parameters values, we calculated the growth and growth ratio of each parameter (Fig.~\ref{fig:Geometrical-comparison-of} and Table~\ref{tab:Value of growth and growth ratio for each parameters}). 

Depression I grew during the perihelion passage. Its length increased by 70\% (13~m), and its width by 50\% (6.5~m) to 140\% (10.1~m) depending on the location inside the depression. The closer to the cliff, the higher  the width growth ratio, indicating that the most significant changes occurred close to this location. The surface area increased by 90\% (181~m$^{2}$) (Fig.~\ref{fig:Geometrical-comparison-of} and Table~\ref{tab:Value of growth and growth ratio for each parameters}). 

Depression II grew by 5~m in length (20\%), and by 5~m, 7~m, and 6~m (2--10\%) at 25\%, 50\%, and 75\% of the length. The longitudinal growth is consistent with the morphological observations
and explains the alteration of the hook shape.

\begin{table*}
\caption{\label{tab:Value of growth and growth ratio for each parameters}Value of growth and growth ratio for each parameter.}      
\centering          
\begin{tabular}  {c c c c c c c c c} 
\hline\hline      
Parameter & Length & Width & Width& Width & Area &  & {Volume (m$^3$)}  &\\ 
 &  (m)  &  25\% (m) &  50\% (m) &   75\% (m) &   (m$^2$) &  Min &   Max & Mean \\ 

\hline\hline
Depression I & & & & & & & &  \\
\hline
\multirow{2}{*}{Growth} & 13.6  & 6.5  & 7 & 10.1 &181&180&2250&1200 \\  
    &  ($\pm \ 1.04$)&($\pm \ 1.04$)&($\pm \ 1.04$)&($\pm \ 1.04$)&($\pm \ 22$)&($\pm \ 23$)&($\pm \ 289$)&($\pm \ 153$)\\
\multirow{2}{*}{Growth ratio} & 1.7 & 1.5 & 1.5 & 2.4 & 1.9 & 1.9 & 3.6 & 3.2 \\
   &  ($\pm \ 0.1$)&($\pm \ 0.1$)&($\pm \ 0.1$)&($\pm \ 0.2$)&($\pm \ 0.2$)&($\pm \ 0.2$)&($\pm \ 0.5$)&($\pm \ 0.3$)\\
\hline  
Depression II& & & & & & & &\\
\hline
\multirow{2}{*}{Growth} & 5.2  & 4.6  & 7.2 &6 & 369 &1830&5420&3630 \\  
    &  ($\pm \ 1.02$)&($\pm \ 1.02$)&($\pm \ 1.02$)&($\pm \ 1.02$)&($\pm \ 17$)&($\pm \ 84$)&($\pm \ 271$)&($\pm \ 181$)\\
\multirow{2}{*}{Growth ratio} & 1.21 & 1.06 & 1.1 & 1.08 & 1.15 & 1.20 & 1.31 & 1.27 \\
   &  ($\pm \ 0.05$)&($\pm \ 0.01$)&($\pm \ 0.01$)&($\pm \ 0.01$)&($\pm \ 0.05$)&($\pm \ 0.05$)&($\pm \ 0.06$)&($\pm \ 0.05$)\\
 \hline
    
\end{tabular}
\end{table*}

The results indicate that both depressions grew during the
perihelion passage. For a more detailed analysis, we computed and compared the length-to-width ratio
of the two depressions, before and after perihelion (Table \ref{tab:Length-to-width}). 
For depression II the mean length-to-width ratio amounts to 0.3$\pm$0.1 before perihelion (i.e., depression II is wider than longer, as already mentioned). After perihelion, this ratio amounts to 0.4$\pm$0.1, which means it has not changed within the error bar. Moreover, this ratio is constant for all widths (at 25\%, 50\%, and 75\% of the length), which demonstrates a homogeneous growth of depression II in all directions during the perihelion passage.
For depression I the mean length-to-width ratio pre-perihelion (1.80$\pm$0.14) and post-perihelion (1.71$\pm$0.06) are also identical, within the error bars, but not the value at the different widths. The length-to-width ratio at 75\% has the most significant variation, from 2.47$\pm$0.27 before perihelion to 1.81$\pm$0.08 after perihelion. This point further reinforces our previous conclusion that the greatest growth occurred close to the cliff. 


\begin{table*}
\caption{\label{tab:Length-to-width}Length-to-width ratio (L/W) of the two depressions, calculated before and after perihelion.}      
\centering          
\begin{tabular}  {c c c c c c} 
\hline\hline      
Aspect ratio & \multicolumn{2}{c}{Depression I} &  \multicolumn{2}{c}{Depression II}  \\ 
 L/W&  Before perihelion &  After perihelion &   Before perihelion &  After perihelion \\ 

\hline\hline
25\% & 1.61 ($\pm \ 0.12$)  & 1.79 ($\pm \ 0.08$)  & 0.33 ($\pm \ 0.01$) & 0.37 ($\pm \ 0.01$) \\ 
50\% & 1.33 ($\pm \ 0.10$)  & 1.54 ($\pm \ 0.07$)  & 0.33 ($\pm \ 0.01$) & 0.37 ($\pm \ 0.01$) \\ 
75\% & 2.47 ($\pm \ 0.27$)  & 1.81 ($\pm \ 0.08$)  & 0.33 ($\pm \ 0.01$) & 0.37 ($\pm \ 0.01$) \\ 
Mean & 1.80 ($\pm \ 0.14$)  & 1.71 ($\pm \ 0.06$)  & 0.33 ($\pm \ 0.01$) & 0.37 ($\pm \ 0.01$) \\ 

 \hline
    
\end{tabular}
\end{table*}

\subsection{Volume comparison }

From Fig.~\ref{fig:Geometrical-comparison-of} and Table~\ref{tab:Value of growth and growth ratio for each parameters} we see that the volume of depression I increased by 1200~m$^{3}$ during the perihelion passage, which corresponds to an increase of 220\% compared to the value before perihelion. For depression II, it increased by 3630~m$^{3}$ or 27\%. The changes in depression I are therefore more pronounced than those  in depression II, with more material removed.



For depression I we estimated the total volume of the new boulders which appeared at the  foot of the associated cliff. We identified ten boulders, with diameters in the range 2.4--6.6~m (Fig.~\ref{fig:Erosional-deposit-comparison}a). The total volume of these boulders amounts to 430~m$^{3}$. This volume is underestimated because (i) smaller boulders are also visible, but their size could not be estimated due to the limited spatial resolution of the images (0.55~m/pixel before perihelion and 0.26~m/pixel after perihelion), and (ii) more than 50\% of the erodible material could have been lost during the perihelion passage \citep{Keller_2015_Insolation_erosion}. Nevertheless, while our volume of 430~m$^{3}$ is underestimated, its order of magnitude is only a factor 3 lower than the mean value of 1200~m$^{3}$ (range 180--2250~m$^{3}$) for the amount of material lost in depression I during perihelion. This strongly supports the idea that the new boulders are indeed deposits resulting from the erosion of depression I.

We  also estimated the volume of the deposit observed at the bottom of depression II as  29900~m$^{3}$. This volume was calculated from the topographic profile shown in Fig.~\ref{fig:Erosional-deposit-comparison}d. However, the lack of information on the  shape and depth of the deposit makes the volume estimate uncertain, preventing us from drawing any conclusion on its origin.

\subsection{Interpretation of the comparison: Landslides and erosional deposits}

Before perihelion, depression I and depression II have a mean slope of
37$^\circ$ and 23$^\circ$, respectively (see Figs.~\ref{fig:Morphometric-measrument-for}b and
\ref{fig:7}b-d). 
These values are in the range of the ``intermediate-slope terrains'' (20$^\circ$-45$^\circ$), as defined by \citet{Groussin_2015_gravitationnal_slopes_b}.
Since the two depressions are localized at the edge of a cliff,
this intermediate slope facilitated the destabilization of the terrain,
which led to the downslope motion of rocks during the last perihelion passage, explaining
the appearance of new boulders.

\begin{figure*}
\centering{\includegraphics[scale=0.58]{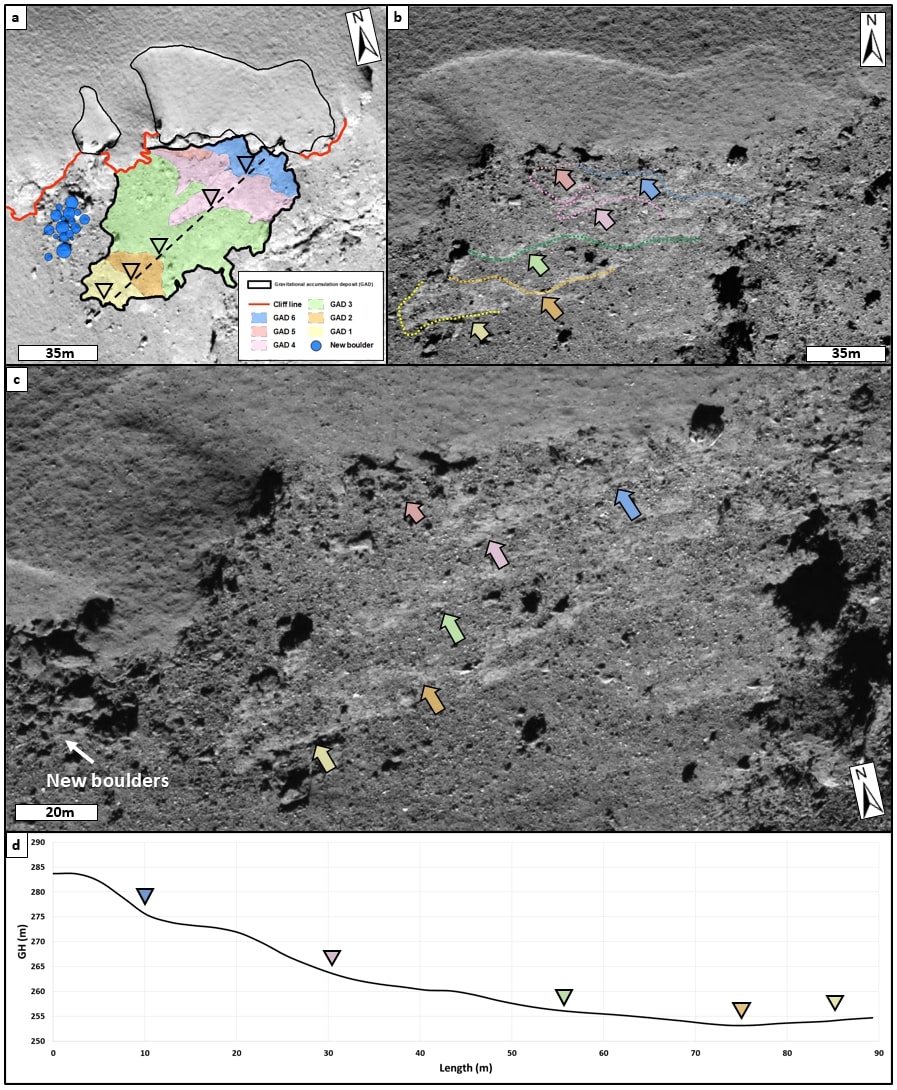}}
\caption{\label{fig:Erosional-deposit-comparison}
Images showing the gravitational deposits associated with depressions I and II.
(a) Image showing the new boulders from depression II, and the different gravitational accumulated deposits (GAD) from depression I likely resulting from successive landslides (see text for details).
(b, c) Images showing the front (colored lines and arrows) of the successive GAD areas.
(d) Topographic profile along the GDA of depression II, corresponding to the dashed line in panel (a). Triangles indicate different positions along the profile, as indicated in panel (a).}
\end{figure*}

We therefore interpret the growth of depression I and its associated deposit as triggered by a landslide event, according to the definition of \citet{Landslide_book}.
For depression II, the lack of new deposits after perihelion does not allow us to easily draw the same conclusion.
However, the characteristics of the observed deposits can help us to constrain its origin.
 \citet{Giacomini_2016_global_north_map}  define this textured material, composed of boulders covered by FPD, as gravitational accumulation deposits, located downslope of depression II.
Those deposits were already present before perihelion, and the fact that the boulders are more degraded (smaller, fragmented, covered by FDP, and/or eroded by sublimation; \citealt{Pajola_2015}) compared to the new ones of depression I, is a strong indication that they are most likely former deposits resulting from an ancient downslope motion of rocks. 
We therefore propose that depression II is an older landslide that was not  activated during the last perihelion passage (Fig.~\ref{fig:Erosional-deposit-comparison}). Nevertheless, because depression II grew, we infer that another erosion process, which has to be different from rock motion (i.e., not a landslide), occurred during the last perihelion passage (see section 5.1).

Finally, the deposits associated with  depression II reveal a stair-like topography (Fig.~\ref{fig:Erosional-deposit-comparison}d). From the images, this deposit seems to cover the surrounding FDP (Fig.~\ref{fig:NAC-images-of}c; Fig.~\ref{fig:Erosional-deposit-comparison}b,c). From the DTM, we identified at least three steps, separated by a slope exceeding 25$^{\circ}$.  Each step is 5 to 10 m in height, and has an increasing length from upstream to downstream. 
We propose three ideas to explain this peculiar topography:
\begin{itemize}
  \item The gravitational accumulation deposits come from a single massive landslide event, which covered a pre-existing stair topography.
  \item The gravitational accumulation deposits also come from a single massive landslide event, which was then followed by several successive collapses that formed the stair topography. 
  \item The gravitational accumulation deposits come from multiple successive landslides, which occurred during different previous perihelion passages, resulting in a stair topography.
\end{itemize}
Due to the regular surrounding topography, the scenario of a unique landslide covering a pre-existing stair topography is not favored. We therefore prefer the third scenario that hypothesizes several successive landslides.

\section{Dynamical evolution of the surface}

\subsection{Erosion processes}

In order to identify the type of landslide for depressions  I and II, we compared their properties with Earth's landslides. Landslides on Earth  are classified depending on the type of movement and material involved \citep{Landslide_book}. 
There are five types of movement, defining five families of landslide: fall, topple, slide, spread, and flow. 
It is also possible to have a complex landslide, which either combines two or more of these types, or which does not fit  any of them \citep{Hungr_2013}.

On the studied depressions we observed movements of rocks detaching from a cliff, without size sorting or visible boulders tracks. These movements are either recent, inducing new boulders for depression I, or older with gravitational deposit accumulation for depression II. These landslides seem to be controlled by gravity, ice sublimation, or both. Therefore, we can exclude a slide (only valid for a coherent mass collapse), a spread (only valid for very gentle slopes), and a flow (only valid in the presence of a liquid phase); we therefore remain with only two types, fall and topple. 

On the one hand, a rock fall is a gravitational effect defined by the detachment of rocks from a steep slope along a surface where little or no shear displacement occurs. The collapsed material falls, bounces, and/or rolls downslope. On the other hand, a block topple is characterized by a forward rotation out of a slope of a mass of rocks around a point below the center of gravity of the displaced mass. Toppling can be driven by gravity or by the presence of ice in cracks inside the mass \citep{Landslide_book,Hungr_2013}. In both cases we should be able to observe the boulders tracks, but we do not. Two hypotheses could explain the lack of tracks. Either the boulder tracks were present at an earlier stage of the process, but were   covered by FPD or erased by successive rock motions, or the boulder tracks never existed. In the depression I landslide it seems that the fragments interact with each other suggesting a movement of broken bedrock in a flow-like manner, such as a dry flow-like debris avalanche \citep{Hungr_2013,Lucchetti_2019_landslide}.

To summarize, in our study the depressions share the properties of topple and fall landslides, and we cannot distinguish between them. However, they are  characterized by a massive (>430 m$^{3}$), dry flow-like motion of fragmented rocks. This therefore corresponds to a complex type of landslide,  specifically  a rock avalanche driven by a fall or a topple movement.   

In any case the landslide is not the only process that modified the depression edges. Even if we did not observe new deposits for depression II, its edges have been extended. We therefore propose that insolation around perihelion led to the propagation of a sublimation front in the FPD material. This process is predicted by thermal models of cometary nuclei \citep{Fanale_1990,Keller_2015_Insolation_erosion}. This sublimation front would trigger  the growth of depression II and the reshaping of its edges, and the relatively large growth of depression I in the northern direction opposite to the cliff (see Fig.~\ref{fig:Morphological-comparison-of}). If our hypothesis is correct, the shape of both depressions would be controlled by this single process occurring around perihelion. The progression of this sublimation front would also destabilize the surface material and lead to the formation of landslide (rock fall and/or topple), as observed. These short-lived events, sublimation front retreat and landslides induced by thermal stress, could drive the global cliff retreat that is a long-lasting global erosional process \citep{Groussin_2015_Geomorpho_Imhotep,Pajola_2016_aswan,El-Maarry_2017,Hu_2017,Attree2018}.

\subsection{Deposit formation}

While depression II grew isometrically without new visible
erosion deposits, depression I had a preferential growth direction aligned with 
new boulders (Fig.~\ref{fig:Morphological-comparison-of} and Table~\ref{tab:Value of each measured parameter}; see also
Fig.~\ref{fig:Geometrical-comparison-of} and Table~\ref{tab:Value of growth and growth ratio for each parameters}). 
According to our hypothesis,
the process described in section 5.1 would destabilize the surface of 
depression I, and the gravitational acceleration would lead to a collapse of boulders at the
bottom of the cliff (Fig.~\ref{fig:Erosional-deposit-comparison}a). 
This collapse is consistent with the morphological evolution of depression I, with a 
larger growth in the length direction close to the cliff edge (Fig.~\ref{fig:Geometrical-comparison-of} and Table~\ref{tab:Value of growth and growth ratio for each parameters}). Even if we did not observed new deposits for  depression II, we can assume that the gravitational accumulation deposit is composed of the same material as the boulders of depression I, but older. We propose that erosion and/or particle deposits have modified the older deposits \citep{Thomas_2015_Redistribution_of_particles}. 

We interpret this gravitational accumulation deposit as the result of one or several previous landslides.
This deposit seems to be superimposed and presents distinct fronts on an intermediate-slope terrain (see Fig.~\ref{fig:Erosional-deposit-comparison}b-d). 
A possible scenario explaining the presence of these fronts would be the occurrence of several 
episodes of cliff collapse during one or more perihelion passages.
If this scenario is true, our observations suggest that even if we did not observe a new
deposit for depression II, one or more previous landslides triggered by sublimation occurred and  progressively shaped depression II and its associated deposits. Another scenario could be that the cliff collapse(s) that formed the gravitational accumulation deposits are not linked with the previous growth of depression II. We could indeed assume that the cliff collapse episode(s) that created the gravitational accumulation deposits occurred before the formation of depression II. Thus, depression II would post-date the gravitational accumulation deposits. The fact that the cliff collapse and the growth of depression I occurred at the same perihelion passage is a strong argument to indicate that these two phenomena are linked, and  we therefore favor the first hypothesis. 

Finally we found that the length of the deposits is decreasing with time,
each new deposit being smaller than the previous one (see Fig.~\ref{fig:Erosional-deposit-comparison}b).
Three hypotheses could explain this observation, although we do not determine which is the most favorable: 
\begin{itemize}
 \item The quantity of erodible material in depression II decreases from one cliff collapse 
 to the next, leading to a decrease in the mass associated with each new deposit.
 \item Following each cliff collapse the deposited material accumulates and levels off the terrain at the foot of the cliff,
 decreasing the gravitational slope. As a consequence, the newly eroded material does not propagate as far as the previous ones.
 \item The variability of the deposit length is controlled by the fraction of ice present in the 
 subsurface layers \citep{Lucchetti_2019_landslide}. If the deposit is induced by several perihelion
 passages, the ice fraction may decrease with time and thus reduce the length of each new deposit.
\end{itemize}

\subsection{Possible chronology}

\begin{figure*}[!ht]
\centering{\includegraphics[scale=0.38]{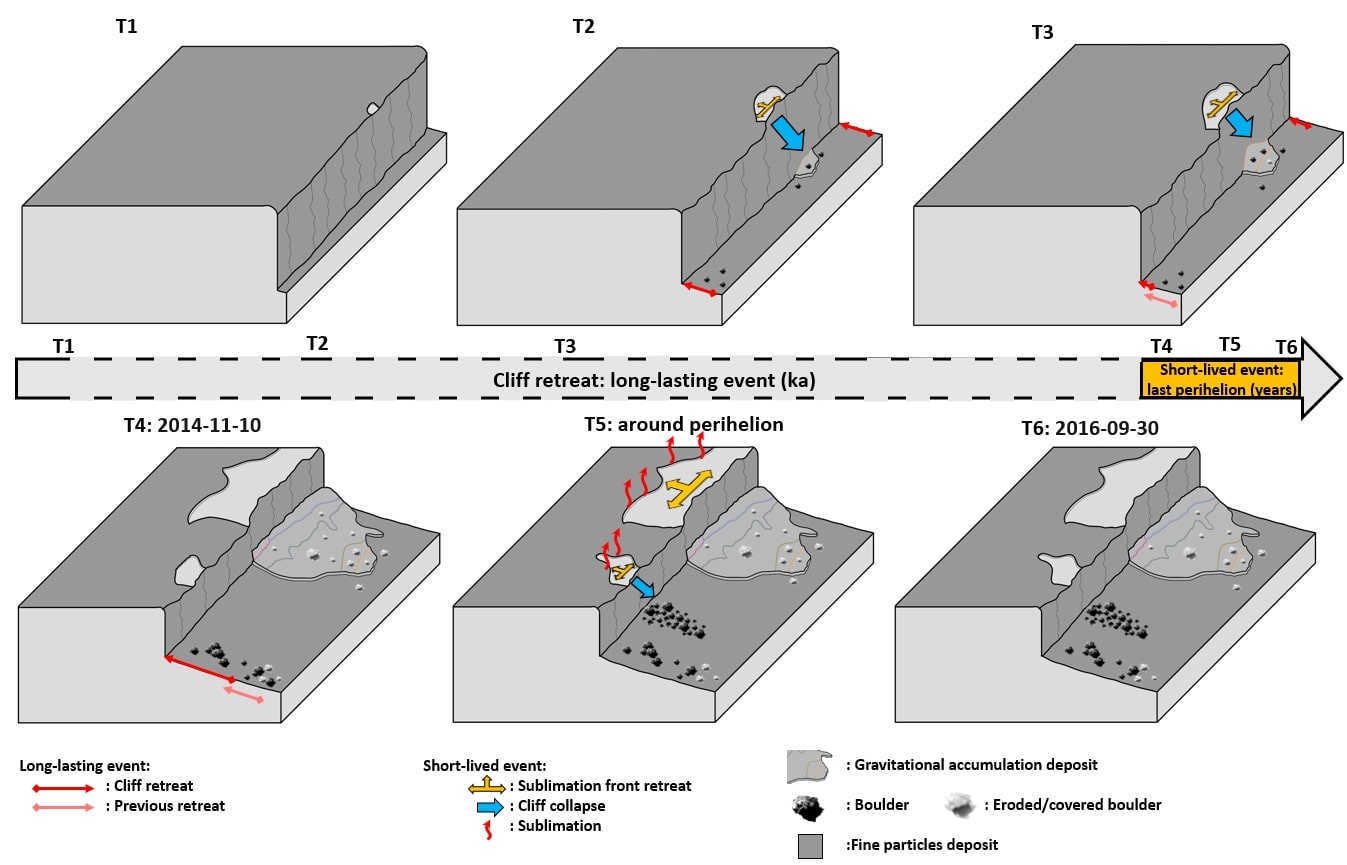}}

\caption{\label{fig:Possible-scenario-for}Possible scenario for the evolution of the two depressions studied in the Ash region (see text for details).}
\end{figure*}

Cliff collapse is an important process that reshapes the nucleus surface of a comet
\citep{Britt_2004,Pajola_2015,Pajola_2016_aswan,Pajola_2017,Steckloff_2016,Steckloff_2018,Lucchetti_2019_landslide}.
The morphology of the associated deposits is controlled by different
factors such as the topography, the slope, the mechanical properties
of the collapsing material, the presence of volatiles, and the insolation
conditions \citep{Cruden_1996,Martha_2010,Hungr_2013}. 
These collapses
are possible due to the rocky nature of the 67P landslides, as
highlighted by \citet{Lucchetti_2019_landslide}. 

From our results we propose a scenario for the formation and evolution of the depressions and landslides observed on Ash, starting from their (unobserved) initial state to what we see today (Fig.~\ref{fig:Possible-scenario-for}). Our scenario takes into account different timescales, denoted  T1 -- T6. First, long-lasting events (ka) occur, such as the cliff overall recession (T1 to T4).Then, during these long and continuous events, short-lived events (months to years) regularly happen. The depression growth and landslide, observed during the Rosetta mission and in this study, reshape the surface at each perihelion passage (T4 to T6). The timescales  are described below: 

T1: Initial state. The cliff is regular and higher than today. The surface is covered by fine particles deposits and only shaped by a small  ``paleo-depression II''. The surface refractory material hides the underlying ice-rich layers \citep{Fornasier_2016}.

T2: Cliff recession and depression grow. At this stage each perihelion passage triggers the sublimation of ices, which causes the growth of   paleo-depression II and a cliff retreat of several meters over time. The growth of   paleo-depression II is a short-time event compared to the cliff retreat. These events lead to the first landslide, and consequently the appearance of the first gravitational deposit and associated boulders.

T3: Similar to T2. The repeated perihelion passages cause the additional growth of  paleo-depression II and its associated gravitational deposits, and the cliff retreat. The boulders are  eroded and/or covered by FDP. 

T4: After many perihelion passages, depression II and its associated deposits have now reached this state, as observed on 23 Jan 2015. The cliff retreated by tens of meters compared to the initial state, and this is the birth of depression I with its associated boulder deposits (Fig.~\ref{fig:Morphological-comparison-of}a; Table~\ref{tab:Value of each measured parameter}). The boulders created during previous perihelion passages are now eroded, and are comprised in a fine material-matrix that formed the GAD observed on 23 Jan 2015.

T5: Around the last perihelion passage, which occurred on 13 Aug 2015, the sublimation of ices lead to the growth of depressions I and II, and to a landslide in depression I (fall or topple) with the appearance of new boulder deposits. 

T6: Last observed state, on 30 September 2016. Beyond T6, for future perihelion passages, due to the proximity of the two depressions, it is possible that they will join each other to form a single, larger depression. The overall cliff retreat should also continue. 

We note that the two erosion mechanisms observed here,  cliff collapse/landslide and  retreat of the sublimation front (depression growth), are not always associated. 
Most of the results published to date only report cliff collapses triggered by sublimation and/or thermal fracturing \citep{Birch_2017_geomorph67P,El-Maarry_2017,El-Maarry_2019,Lucchetti_2017,Lucchetti_2019_landslide,Pajola_2016_aswan,Pajola_2017,Pajola_2019}, and do not detect any associated depression growth. We will therefore continue our analysis on more depressions, all over the nucleus, to better understand the combination of these two erosion processes and to further constrain our scenario.

\section{Conclusion}

We used OSIRIS images of the nucleus of comet 67/Churyumov-Gerasimenko acquired by the Rosetta spacecraft to make a comparative quantitative
morphometrical analysis of two depressions located in the Ash
region. Our approach combines the analysis of high-resolution images from
the NAC with geographic information products (GIS) and, in particular, the  topography from the SPC  shape model. 

We observed new temporal morphological changes at the surface of the Ash region,
and we quantified those changes on the metric scale.
Our morphometrical analysis revealed that the two studied depressions (I and II)
grew by several meters compared to the last perihelion passage; however, 
this growth is not identical in the two depressions. 
While depression II grew in all directions, depression I grew preferentially in the part located close to the cliff.
Moreover, while depression II does not show evidence of new deposits, the growth of depression I close to the cliff created new boulder deposits at
the foot of the  cliff. 
The two depressions show evidence of a complex landslide (rock avalanche), which has either a topple or a rock fall movement. 
For depression II we observe an ancient gravitational accumulation of
deposits at its foot, which indicates that this depression
has been shaped by several perihelion passages. The sublimation of ices certainly played a key role in shaping this overall area.

Finally, we propose a qualitative chronology for the formation and evolution of the two depressions. Our chronology raises the   question of   erosion by sublimation of ices on long timescales for the overall cliff retreat versus short timescales for depression growth and landslides. Additional evidence of similar events in other regions of the nucleus of 67P are certainly required to strengthen this scenario and to better constrain the erosion processes. 

\begin{acknowledgements} OSIRIS was built by a consortium of the
Max-Planck-Institut fur Sonnensystemforschung, Katlenburg-Lindau,
Germany; CISAS University of Padova, Italy; the Laboratoire d'Astrophysique
de Marseille, France; the Instituto de Astrofisica de Andalucia, CSIC,
Granada, Spain; the Research and Scientific Support Department of
the ESA, Noordwijk, Netherlands; the Instituto Nacional de Tecnica
Aeroespacial, Madrid, Spain; the Universidad Politechnica de Madrid,
Spain; the Department of Physics and Astronomy of Uppsala University,
Sweden; and the Institut fur Datentechnik und Kommunikationsnetze
der Technischen Universitat Braunschweig, Germany. The support of
the national funding agencies of Germany (DLR), France (CNES), Italy
(ASI), Spain (MEC), Sweden (SNSB), and the ESA Technical Directorate
is gratefully acknowledged. We thank the Rosetta Science Operations
Centre and the Rosetta Mission Operations Centre for the successful
rendezvous with comet 67P/Churyumov-Gerasimenko. The authors are very grateful to the anonymous reviewer for his/her very constructive
comments and suggestions that improved this paper
\end{acknowledgements}

\bibliographystyle{aa}
\bibliography{biblio_bibtex}

\end{document}